\documentclass[aps,prl,twocolumn,showpacs,floatfix,amsfonts,amssymb]{revtex4}

\usepackage{graphicx,graphics,color,epsfig}
\usepackage{bm}
\usepackage{amsmath}
\usepackage{amssymb}
\usepackage{amsfonts}

\newcommand{\beqa}{\begin{eqnarray}}
\newcommand{\eeqa}{\end{eqnarray}}

\newcommand{\HeThree}{$^3$He}
\newcommand{\HeFour}{$^4$He}
\begin{document}

\title{Entropy of solid \HeFour: the possible role of a dislocation glass}

\author{A. V. Balatsky$^{1 }$, M. J. Graf$^{1}$, Z. Nussinov$^{2}$,
and S. A. Trugman$^{1}$ }
\affiliation{$^1$Theoretical Division, Los Alamos National Laboratory, Los Alamos, New Mexico 87545, USA}
\affiliation{$^2$ Department of Physics, Washington University, St. Louis, MO 63160 USA}
\email{avb@lanl.gov}
\date{\today}

\begin{abstract}
Solid \HeFour\ is viewed as a nearly perfect Debye solid. Yet,
recent calorimetry indicates that its low-temperature specific
heat has both cubic and linear contributions. These features appear in the same
temperature range ($T \sim 200$ mK) where measurements of the torsional oscillator
period suggest a supersolid transition. We analyze the
specific heat to compare the measured with the
estimated entropy for a proposed supersolid transition with 1\%
superfluid fraction. We find that the experimental
entropy is substantially less than the calculated entropy.
We suggest that the low-temperature linear term
in the specific heat is due to a glassy state that develops at low
temperatures and is caused by a distribution of tunneling systems in
the crystal. It is proposed that small scale dislocation loops produce those tunneling systems.
We argue that the reported  mass decoupling is consistent with an
increase in the oscillator frequency 
as expected for a glass-like transition.
\end{abstract}

\maketitle

\paragraph{Introduction.}

A supersolid is a novel state of matter that simultaneously
displays both superfluidity
and crystalline order. \HeFour\ is thought to be
a most likely candidate for the supersolid state. Recent
torsional oscillator experiments by Kim and Chan
\cite{Chan04,Chan:05} generated renewed interest
in this possibility. Pioneering
work by Andreev and Lifshitz \cite{AL}, Reatto
\cite{Reatto67}, Chester \cite{Chester70}, Leggett \cite{Leggett70},
and Anderson \cite{Anderson84} laid the foundation for our
thinking about this enigmatic state. Recently, Anderson et al.
revisited this problem \cite{Anderson05,ABH} and latest 
developments, presented at a KITP workshop, are available online \cite{KITP}.

In addition to the work of the PSU group \cite{Chan04,Chan:05},
there is now an independent confirmation of the anomalous behavior
of solid \HeFour, as presented by the groups of Reppy and
Shirahama\cite{Sophie06,Shirahama06}. Both groups use torsional
oscillators similar to the one by the PSU group of Chan
\cite{Chan04,Chan:05}. At the same time  Rittner and Reppy
\cite{Sophie06} reported a history dependence of the signal, when
annealing the sample, to the extent of no observation of any mass
decoupling in the torsional oscillator experiment. These torsional
oscillator experiments
 clearly indicate anomalous mechanical properties of solid \HeFour\ at low
temperatures. However, the relationship between the mechanical
measurements and the suggested  superfluidity is not direct. The most
direct proof of superfluidity would be observation of persistent
current. In this regard, we mention a recent experimental search for
superflow by Beamish et. al. \cite{Beamish05} that indicates no mass
flow of any kind to very high accuracy. Thus, we conclude that the
effect first observed by the PSU group is likely not an intrinsic
property of solid \HeFour, because it depends critically on
the \HeThree\ concentration and shows annealing
dependence \cite{Sophie06}.

Alternatively, we suggest that many of the experimental
facts seem to be consistent with a glass-like behavior of
crystalline \HeFour~ at temperatures $T \le 200$ mK.
For the lack of any definitive experiment, we
assume that the glass in \HeFour~ is formed due to dislocations. In
this Letter we would like to decouple the discussion of the observed
features in  specific heat and torsional oscillator from the
supersolidity. We will focus on two critical features reported to
date by experiments: (1) the small entropy that is seen
experimentally near the transition temperature. The observed entropy is
at almost two orders of magnitude smaller than the expected
entropy of a 1\% supersolid fraction. (2) The observed linear specific
heat in the {\em bosonic} crystal of \HeFour, consistent with a glass.

\paragraph{Entropy analysis.}

\begin{figure}[tb]
\begin{center}
\includegraphics[height=6.9cm,angle=270]{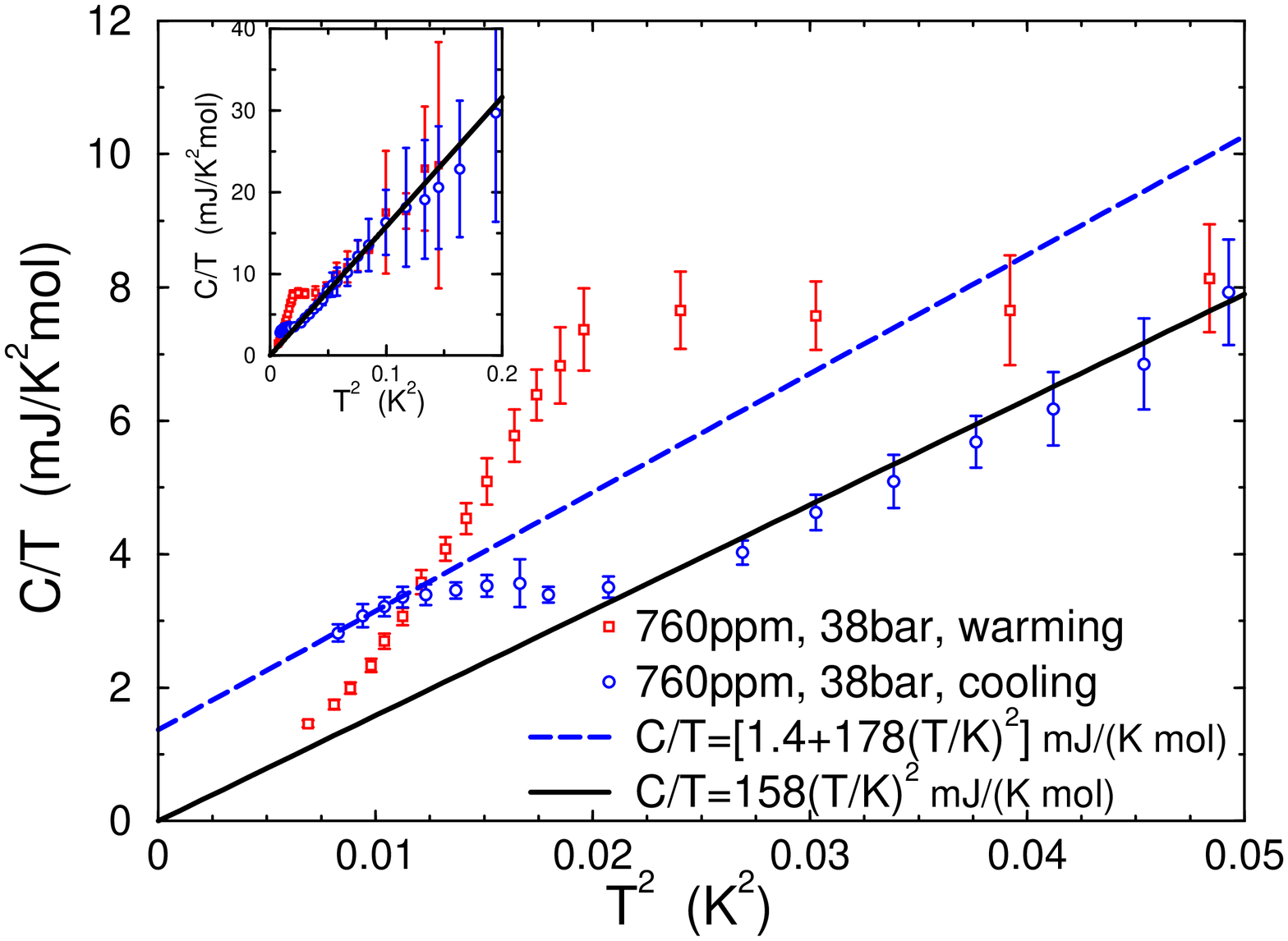}
\includegraphics[height=6.9cm,angle=270]{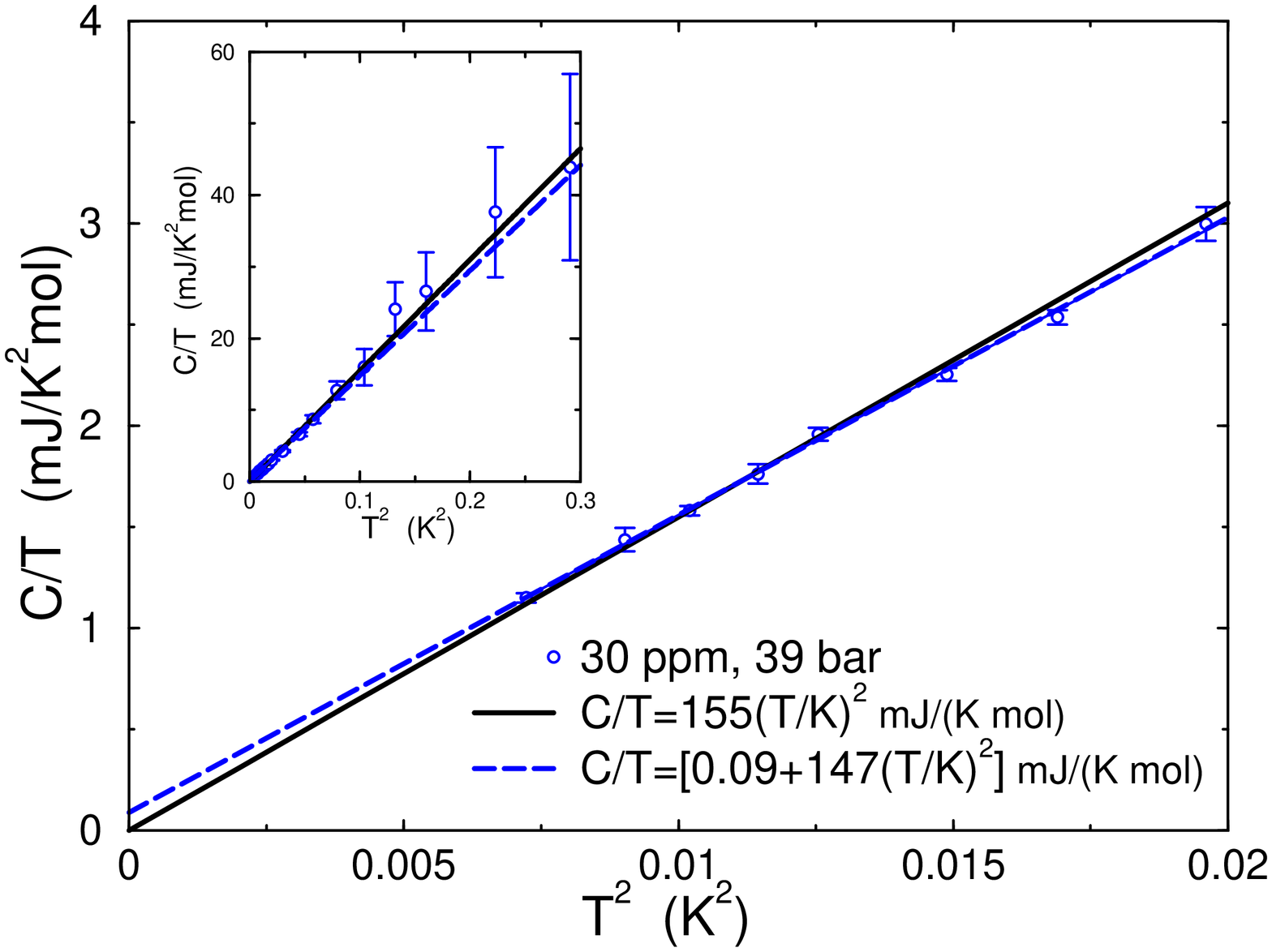}
\end{center}
\caption{Specific heat over temperature of solid \HeFour\ by Clark
and Chan\cite{ChanJLTP04}. (Top): sample with 760 ppm \HeThree\ at
38 bar. (Bottom): sample with 30 ppm \HeThree\ at 39 bar.
The dashed and solid lines are the respective low
and high temperature fits, whose coefficients are presented 
in Table~\ref{table1}. 
The insets show an enlarged temperature region.} \label{FigC}
\end{figure}

We focus on the specific heat measurements on \HeFour~
by Clark and Chan \cite{ChanJLTP04} which indicate a departure from
the conventional $T^3$ specific heat behavior expected at low temperatures.
Given the data, we searched for evidence of a thermodynamic phase
transition to a supersolid phase, assuming a 1\% concentration of
condensate.
The observed features in the specific heat occur in the same
temperature range where a change in the period of the torsional
oscillator led to the speculation of a transition to a supersolid
state (see Fig.~\ref{FigC}). 
A low-temperature linear term in the specific heat can be discerned. 
We note that a linear term in the
specific heat of \HeThree\ and \HeFour\ crystals was observed more
than 40 years ago by Heltemes and Swenson \cite{Swenson62} and
Franck \cite{Franck64}, but not in later measurements by Edwards and
Pandorf \cite{Pandorf65}. A theoretical explanation of the earlier results
was given in terms of thermal vibrations of pinned dislocation
segments based on the theory of Granato \cite{Granato58}. 

\begin{figure}[th]
\begin{center}
\includegraphics[height=6.9cm,angle=270]{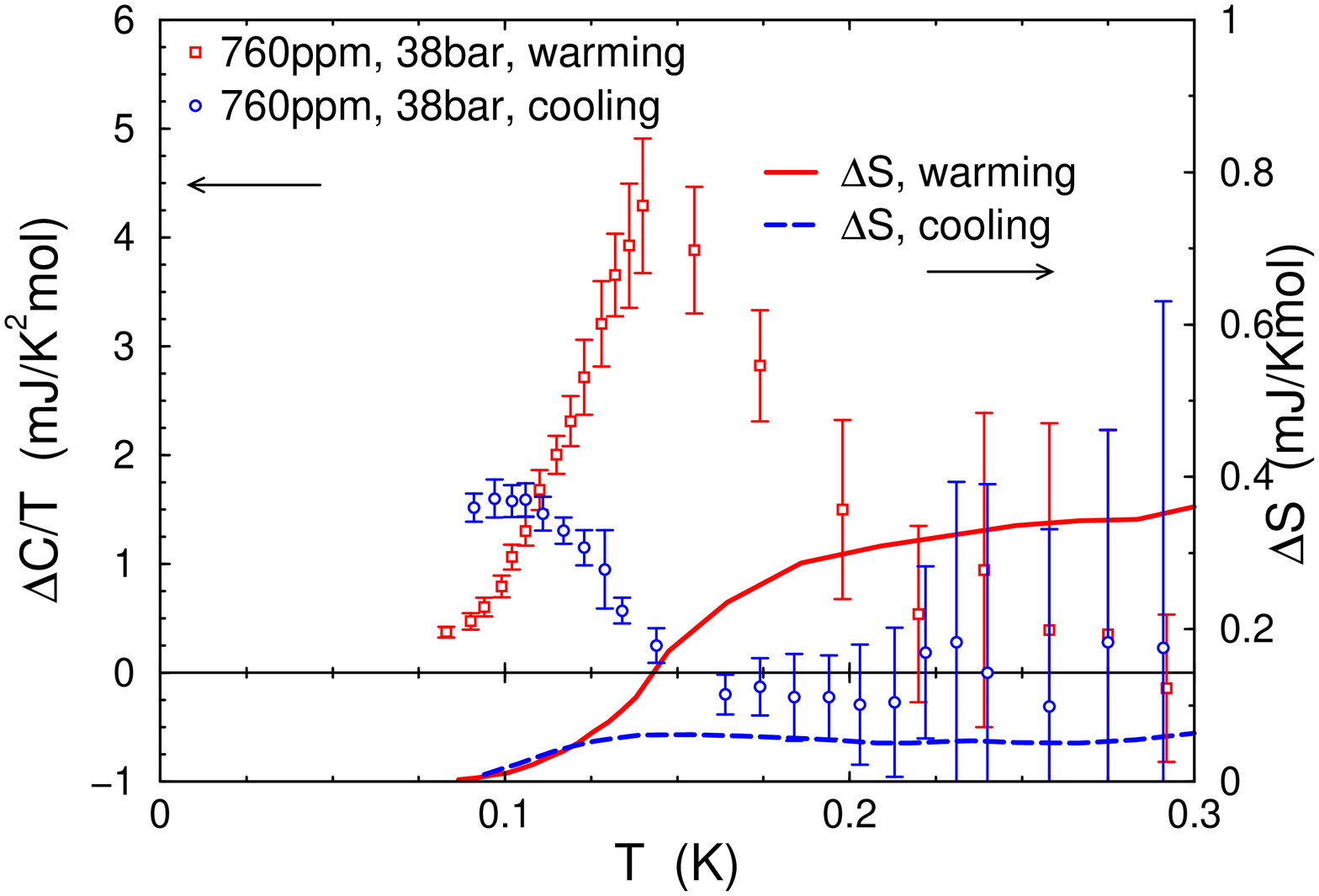}
\includegraphics[height=6.9cm,angle=270]{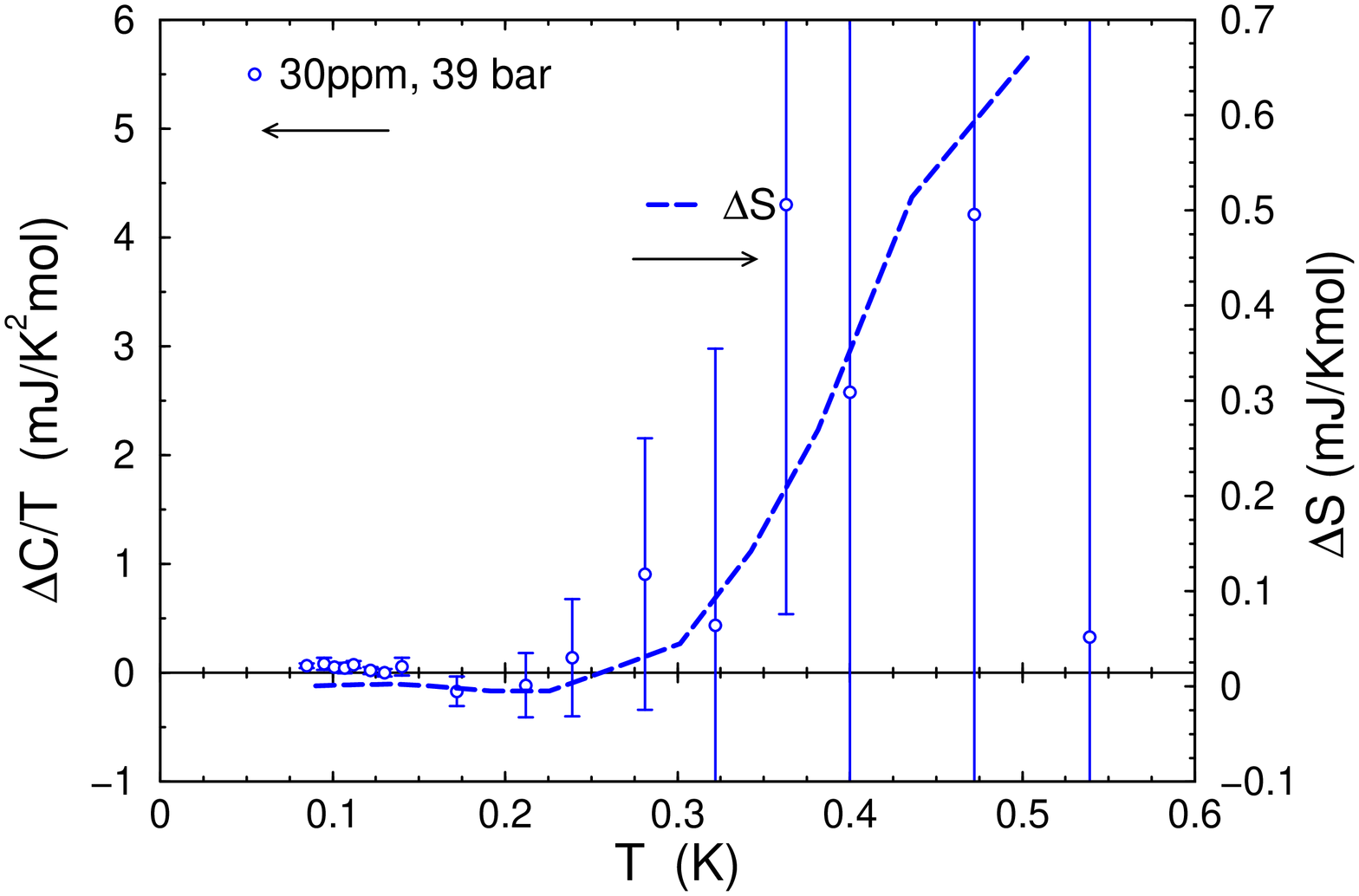}
\end{center}
\caption{The $\Delta C/T$ differences and the corresponding entropies 
$\Delta S = \int_{T_{min}}^T dT' \Delta C/T'$ at low $T$.
(Top): sample with 760 ppm of \HeThree\ at 38 bar for cooling and warming runs; 
the high-$T$ ($T > 200$ mK) contribution of solid \HeFour, $C = B\, T^3$,
with $B_{760 \rm ppm}=158$ mJ/(K$^4$ mol) was subtracted. (Bottom):
sample with 30 ppm of \HeThree\ at 39 bar; the high-$T$
contribution with $B_{30 \rm ppm}=155$ mJ/(K$^4$ mol) was subtracted. 
}
\label{FigEntropy}
\end{figure}

We now perform a general thermodynamic analysis of the
measured specific heat data by Clark and Chan (see
Figs.~\ref{FigC} and \ref{FigEntropy}).  This is especially attractive for the analysis
as its result will not depend sensitively on the exact functional
form of the specific heat $C(T)$ at low temperatures, $T \leq 100$
mK. Most of the change in the entropy, $\int^{T} dT'
\frac{C(T')}{T'}$, which we find from the data, originates from the
region of 100-400 mK. According to the torsion oscillator
measurements the supposed solid to supersolid transition occurs in
the temperature range 160-320 mK. The low-$T$ entropy
differences of the 760 ppm and 30 ppm samples relative to solid
\HeFour\ are roughly $\Delta S(760\,\rm ppm) \sim 0.06-0.4$ mJ/(K mol) and
$\Delta S(30\,\rm ppm) \sim 0.3$ mJ/(K mol) respectively, between
80 mK and 400 mK, see Fig.~\ref{FigEntropy} \cite{entropy}.

We assume, consistent with the experimental findings,
that roughly 1\% of the bulk sample
transforms into a superfluid. Thus 99\% of the sample is unaffected and
behaves like a perfect Debye crystal below $\sim 500$ mK ($T < \Theta_D/50$).
Hence, the supersolid fraction would indicate
an entropy of $ S = 46$ mJ/(K mol).
We now compare extracted entropy changes with
what may be expected from a supersolid
to solid transition (i) in a dilute gas, 
as suggested by a 1\% supersolid component,
and (ii) in a strongly correlated dense gas:
(i) The Bose Einstein condensation (BEC) of an ideal gas gives values
of entropy changes which are an order of magnitude larger than
those observed for the $\lambda$ transition
of liquid \HeFour. We follow suite here and compare
the ideal BEC with that
of a possible supersolid transition.
Assuming a noninteracting Bose-Einstein
gas in three dimensions with a parabolic band,
the specific heat of the BEC is 
$C(T) = \frac{15}{4} \frac{\zeta(5/2)}{\zeta(3/2)} R (T/T_{c})^{3/2}$ 
for $T \le T_c$, with $\zeta(z)$ the Riemann zeta function and
gas constant $R$.
The total entropy per mole in the condensed
state taken between $T= 0$ and  $T_c$ is universally
$S_{\rm BEC}(T_c) = \frac{5}{2} \frac{\zeta(5/2)}{\zeta(3/2)} R \approx 5R$.
If we use a  1\% molar concentration of supersolid fraction, 
we still get $S_{\rm BEC}(T_c) \approx 0.05\, R =  416 $ mJ/(K mol) [per mole of
\HeFour]. This is three orders of magnitude larger than the entropy changes
seen in experiments (Fig.~\ref{FigEntropy}). These conclusions do not
change if a short-range repulsive interatomic potential
augments the kinetic energy in a dilute Bose gas.
(ii)  A lower bound can be found by scaling the measured $S_\lambda$
at the $\lambda$-point of strongly correlated superfluid \HeFour\ under pressure
($P_\lambda \sim 26$ bar), which is $S_\lambda = 4.6$ J/(K mol)
at $T_\lambda = 1.8$ K \cite{Ahlers73}. 
As a generous lower bound on the entropy at an actual supersolid
transition, we may correct for the fact that the claimed transition
temperature is 9 times smaller ($\sim$ 200 mK for 30 ppm sample)
\cite{Chan04} than $T_{\lambda}$. We scale down the expected 1\%
entropy by a factor of 9. Then, we may expect to find an entropy of
$S_\lambda \sim 5$ mJ/(K mol). This estimate is about an order of magnitude
bigger than the measured entropies. 
In ideal superfluids a nearly $T_{c}$ independent entropy $S(T_{c})$ results, 
even for different low-$T$ scaling regimes, e.g., phonons, rotons. 
We emphasize that our linear scaling of
$S(T_c)$ is a generous lower bound \cite{clark1.8}.

More sensitive measurements of the specific heat using a new
experimental setup by Chan et. al. (unpublished) are consistent with
the deviations from a $T^3$ behavior, however, the precise form at
lowest temperatures, $T \leq 80$ mK is not settled yet.
Clearly, the estimated entropy associated with an actual superfluid
transition is substantially larger than the measured
$\Delta C/T$ integrated from $\sim 80$ mK  to 500 mK for
the case of $\lambda$ transition or weakly interacting BEC. The
observed deficit of entropy is very hard to reconcile with the
1\% fraction of superfluid transition that is suggested to
account for the torsional oscillator experiments.

A null result of the signature of a $\lambda$ transition,
the observation of hysteresis on cooling and warming and the dependence of the excess
entropy on annealing points to a glass-like phenomenon of
tunneling systems rather than that of a supersolid.

\paragraph{Linear specific heat and sensitivity to \HeThree\ impurities.}

We propose that the linear specific heat term \cite{Chan04,Swenson62,Franck64} 
is due to Tunneling Systems (TS) in the \HeFour\ crystal. Specifically, we
assume that dislocation loops are small enough to be present in large numbers
($\sim 10^{10} {\rm cm^{-2}}$) to  create the TS. We propose that fluctuations of nanoscale
segments of dislocation loops form the TS.
A small addition of \HeThree\ atoms to the \HeFour\ crystal facilitates the
creation of dislocations; hence, the linear specific
heat term should increase with \HeThree\ concentration,
similar to the effect of small amounts of hydrogen in crystalline tantalum \cite{Koenig}.

The distribution of the characteristic energies of the TS is
given by $P(E) = P_0 dE$,
where we follow the standard discussion on the role of two-level
systems in glasses \cite{Anderson72,Phillips72}. It is commonly
assumed that at low energy $P_0$ is constant.
We assume that $P_0$ is only a function of the
\HeThree\ concentration $n_3$ and any contribution from intrinsic
defects in \HeFour\ is neglected. \HeThree~ facilitates the creation of
dislocations and hence $P_0$ should grow with the
concentration of \HeThree. We further
assume that per mole of \HeFour\,,  $P_0(n_3) = c N_A n_3^{\nu}$,
with positive exponent $\nu$ and coefficient $c$, and Avogadro's
number $N_A$. A natural choice is $\nu  = 1$, at least
for low concentrations of \HeThree. To keep the discussion general,
we will not specify $\nu$.

The specific heat of a single tunneling system is
$C_{TS}(E,T)  =
   k_B \left({E}/{k_B T}\right)^2\frac{\exp(E/k_B T)}{(1 + \exp(E/k_B T))^2}$.
The average over the distribution of the TS gives the total molar specific heat,
which at low temperatures is
$C_{TS} = \int_0^\infty dE ~P(E)~ C_{TS}(E,T) \approx {\pi^2 \over 6} k_B^2 P_0 T$. 
The total specific heat of \HeFour\ is the sum of lattice,
($C_{lat}$) and the TS
($C_{TS}$) contributions. For a perfect Debye crystal,
the molar $C_{lat} = \frac{12 \pi^4}{5} R (T/\Theta_D)^3$
at low T.  In solid
\HeFour,  the Debye temperature $\Theta_D \approx 28$ K at $P \approx 40$ bar.
Thus at low temperatures ($T \alt \Theta_D/50 \approx  0.5$ K)
the specific heat per mole of \HeFour\ is
$C = A T + B T^3$,
with $A = \frac{\pi^{2}}{6} k_B^2 P_0$ and 
$B= \frac{12 \pi^4}{5} R / \Theta_D^3$.
For $T< T^* = \sqrt{({\Theta_D^3 k^2_B P_0})/({12 \pi^2}) } \sim n_3^{\nu/2}$,
the linear term dominates over the lattice contribution.
It follows that the TS model predicts that both the
crossover temperature $T^*$ and the linear coefficient $A$
in the specific heat {\em will be very sensitive with respect to \HeThree\ concentration}.
For example, for $\nu = 1$ it leads to a square root dependence on $n_3$. This result
also suggests that the effect of mass decoupling either vanishes or
occurs at a much lower temperatures in samples with vanishing \HeThree\
concentration.

We revisited the data by Clark and Chan \cite{ChanJLTP04}, see
Fig.~\ref{FigC}, and assumed for our analysis linear and cubic terms in
the specific heat. We find that indeed the data are
consistent with a strong dependence on \HeThree\ concentration.  For
example, the linear term depends on \HeThree\
concentration, as shown in Table~\ref{table1}. The $A$ coefficient
for \HeFour\ reported by Heltemes and Swenson\cite{Swenson62} and
Franck\cite{Franck64} varied between $A \sim 2.5 - 8.8$  mJ/(mol
K$^2$) at a similar pressure of $P \approx 40$ bar and with a Debye temperature
$\Theta_D \approx 28$ K.\cite{clark1.8}
Their $A$ coefficients are somewhat bigger but of the same order as ours, 
which are in the range of
$A= 0.09-1.4$ mJ/(K$^2$ mol), depending on \HeThree\ concentration and history.

The extracted low-temperature linear term coefficient $A$ scales roughly
linearly with the \HeThree\ concentration. Allowing for
different functional forms for the specific
heat for various systems of interacting bosons does not lead
to a significant change in our results.

\begin{table}[tb]
\caption{Summary of the linear and cubic coefficients of $C=A T + B T^3$ of
solid \HeFour\ with \HeThree\ solute, as well as its Debye temperatures $\Theta_D$.
The $A$ is from fits below $\sim 120$ mK, while the
$B$ and $\Theta_D$ are from fits between $200$ mK $< T \alt
500$ mK. Rough estimates of the uncertainties are in parentheses.
For 30 ppm, $A$ is close to zero as expected for this concentration.
}\label{table1}
\begin{tabular}{ccccc}
\hline
$^3$He & $P$ & $A$              & $B$            & $\Theta_D$ \\
 ppm   & bar & mJ/(K$^2$ mol)   & mJ/(K$^4$ mol) & K \\
\hline
30      	 & 39    & 0.09(4) & 155(5) & 23.2       \\
760 (cooling)    & 38    & 1.4(2)  & 158(5) & 23.1	  
\end{tabular}
\end{table}

\paragraph{The ``missing'' moment of inertia and susceptibilities.}

The torsional oscillator
experiments measure the susceptibility - they do not
directly monitor the moment of inertia of the supersolid.  As in any time
translationally invariant system,  the Fourier
amplitude of the angular response of the torsion oscillator is
$\theta(\omega) = \chi(\omega) \tau_{ext}(\omega),$
with $\chi = \chi_{1}+ i \chi_{2}$ an angular
susceptibility and $\tau_{ext}$
the external torque. For the simple torsional
oscillator, $\chi^{-1}(\omega,T) =
[\alpha - i \omega \gamma_{osc} - I_{osc} \omega^{2}+ g(\omega,T)]$.
Here, $I_{osc}$ is the moment of inertia of the torsional oscillator,
$\alpha$ is its restoring constant, $\gamma_{osc}$ is the dissipative
coefficient of the oscillator, and $g(\omega,T)$ arises from
the back action of solid \HeFour. For an ideal normal
solid with moment of inertia $I_{ns}$, which rotates
with the oscillator, the back action is
$g= - I_{ns}(T) \omega^{2}$. However, we do not
need to impose this form.
Current experiments measure the oscillator
period  $2 \pi/ \omega_{0}$  with $\omega_{0}$
the real part of the solution of $\chi^{-1}(\omega,T)=0$
at fixed $T$.  For example, a decrease in an
effective dissipative component ($-i \gamma_{glass} \omega$)
in $g(\omega, T)$ as the temperature is lowered
(wherein a liquid component transforms into a solid glass)
will also lead to a shorter rotation period.
The decrease in the rotation period
only implies a crossover in $\chi$ (and a
constraint on $g$). As the real and imaginary parts of $\chi$
are related by the Kramers-Kronig relations,
an enhanced decrease in $\chi_{1}(\omega,T)$
often appears with a pronounced peak
in $\chi_{2}$ \cite{Sophie06}.
A nonvanishing $\chi_{2}$ at finite frequency mandates dissipation.

The microscopic deformation of the glass to the applied torsion
might, similar to suggestions \cite{epl_danna} concerning torsional
oscillator results on granular media \cite{nature_anna}, take the
form of elastic, plastic, and fracture processes (including those of
internal avalanches). A gradual change can proceed through
dislocation glide in the slip plane \cite{hiki}. All these
possibilities need to be addressed experimentally before we can have
a definitive microscopic picture of the possible glass state.

One way to differentiate between glassy effects
and a true thermodynamic $\rho_s$ is to measure the frequency
dependence of $\rho_s$. For the dislocation TS that we propose,
$\rho_s(T, \omega)$ will be a nontrivial function of frequency
which vanishes as a power of frequency $\rho_S(T \rightarrow 0,\omega)
\sim \omega^{\alpha}$. On the other hand,
for a true supersolid phase there is a finite limit $\rho_S(T
\rightarrow 0,\omega) \sim \rho_s(0)$.

\paragraph{Heat Pulse Experiment. }

A heat pulse experiment may test for the existence
and relevance of the TS for the thermal properties \cite{Nittke98}.
In the absence of TS, the heat pulse would trigger quick
equilibration of the energy with the phonon bath. Hence,
on a very short time scale
the temperature of a sample will reach a steady state value. By contrast,
for a crystal with TS, the heat deposited in the crystal will first
be absorbed by the phonon bath and later will cause the re-population
of the TS as a result of the energy transfer from phonon bath to TS.

\paragraph{Conclusions.}

We find that the measured
entropy excess is several orders of magnitude smaller than the
entropy expected from the BEC or $\lambda$ transition of 1\%
superfluid fraction. While the absence of the entropy released at
claimed supersolid transition is puzzling,
it is consistent with a dislocation induced glassy state in \HeFour\
crystals. This hypothesis predicts that the linear term in the
specific heat increases with \HeThree\ concentration.
Heat pulse, heat transport and ultrasound measurements would be helpful in
identifying tunneling systems if they indeed exist in crystalline
\HeFour. The increase in the oscillator frequency
at low temperatures is consistent with this interpretation.

\paragraph{Acknowledgements.}

We acknowledge discussions with E. Abrahams,  P. W.
Anderson, D. Arovas, C. D. Batista, B. Brinkman, M. Chan, D. Ceperly,
A. Dorsey,  D. Huse, J. Goodkind, A. J. Leggett, J. Reppy, and G.
Zimanyi. This work was supported by US DOE. AVB is grateful to
the  KITP for its hospitality during the Supersolid Mini-Program.


\begin{thebibliography}{99}
\bibitem{entropy} 
Instead of integrating $\Delta C / T$, one could alternatively
directly compare the function $\Delta C(T) / T$ to what is expected 
for the BEC.  In Fig.~\ref{FigC} (top, cooling), the
low-temperature bump has nearly the expected form for
the BEC, but is approximately 500 times too small.
\end{thebibliography}
\end{document}